\documentstyle[11pt,twoside,colloqOHP,epsfig]{article}
\begin{document}
\title{NAHUAL: A cool spectrograph for planets of ultra-cool objects}
\author{E.W. Guenther$^1$, E.L. Mart\'\i n $^{2,3}$, D. Barrado y Navascu\'es$^4$, U. Laux $^1$}
\affil{$^1$ Th\"uringer Landessternwarte Tautenburg, Sternwarte 5, 
            D-07778 Tautenburg, Germany[guenther@tls-tautenburg.de]} 
\affil{$^2$ Instituto de Astrof\'\i sica de Canarias, E-38200 La Laguna (Tenerife), Spain} 
\affil{$^3$ University of Central Florida, Dept. of Physics, PO Box 162385, 
            Orlando, FL 32816-2385, USA}
\affil{$^4$ LAEFF - INTA, PO Box 50727, E-28080 Madrid, Spain} 
 

%
\begin{abstract}
We present the status of an ongoing study to built a a high resolution
near infrared Echelle spectrograph (NAHUAL) for the 10.4-m-Gran
Telescopio Canarias (GTC) which will be especially optimised for planet
searches by means of high precision radial velocity measurements.  We
show that infrared radial velocity programs are particularly suitable to
search for planets very low mass stars and brown dwarfs, as well as
active stars. The goal of NAHUAL is to reach an accuracy of the radial
velocity measurement of a few $ms^{-1}$, which would allow the detection
of planets with a few earth-masses orbiting low-mass stars and brown
dwarfs. It is planed that NAHUAL covers simultaneously the full
wavelength range in the J, H, and K-band, and will also serve as a
general purpose high resolution near infrared spectrograph of the GTC.
The planed instrument will have a resolution of $\lambda/ \Delta
\lambda=50,000$ with a 0.175 arcsec slit, and an AO-system. An
absorption cell will serve as a simultaneous wavelength reference.
\end{abstract}
\section{Introduction}

In this contribution, we will present a study for a high-resolution
spectrograph which is especially designed for high precision radial
velocity (RV) measurements at near infrared (IR) wavelength. The
instrument is called NAHUAL \footnote{A NAHUAL is also a kind of shaman
in Mexican mythology that is a person in daytime but a hunter at night
time.}  for Near-infrAred High-resolUtion spectrogrAph for pLanet
hunting, and will be operated at the 10.4-m-GTC telescope (Gran
Telescopio Canarias). The GTC will see first light in 2006. NAHUAL will
also serve as a high-resolution spectrograph IR spectrograph for general
use (Mart\'\i n et al. 2005). In the second section of this
contribution, we will discuss the benefits if RV-planet search programs
are being carried out at IR-wavelength, and in the third and fourth
section, we will discuss the requirements and present the conceptional
design of the instrument.

\section{The benefits of high-resolution NIR spectroscopy for exo-planet research}

\subsection{Planets of brown dwarfs and very low-mass stars}

Most of the efforts for detecting extra-solar planets have hitherow been
concentrated on main sequence F,G,K stars. These surveys show that while
a large fraction of the stars are binaries and many have massive
planets, there is a lack of close-in brown dwarf (BD) companions. This
result has often been used as an argument that there are two distinct
formation tracks: one leading to planets, and the other to stellar
companions.  In the standard model massive planets form by core
accretion: In the first step a solid core of about 0.01 to 0.03
$M_{Jupiter}$ forms, which subsequently accretes gas from the disk in
order to form a massive planet.  The core accretion scenario is
supported by the fact that stars with an overabundance of heavy elements
also have a higher frequency of massive planets (Santos, Israelian, \&
Mayor, 2004). The discovery of HD\,149026\,b, which has 0.21
$M_{Jupiter}$ core composed of elements heavier than helium also
supports the core accretion scenario (Sato et al. 2005).

\begin{figure}[h]
\vspace*{7.0cm}
\begin{center}
\includegraphics{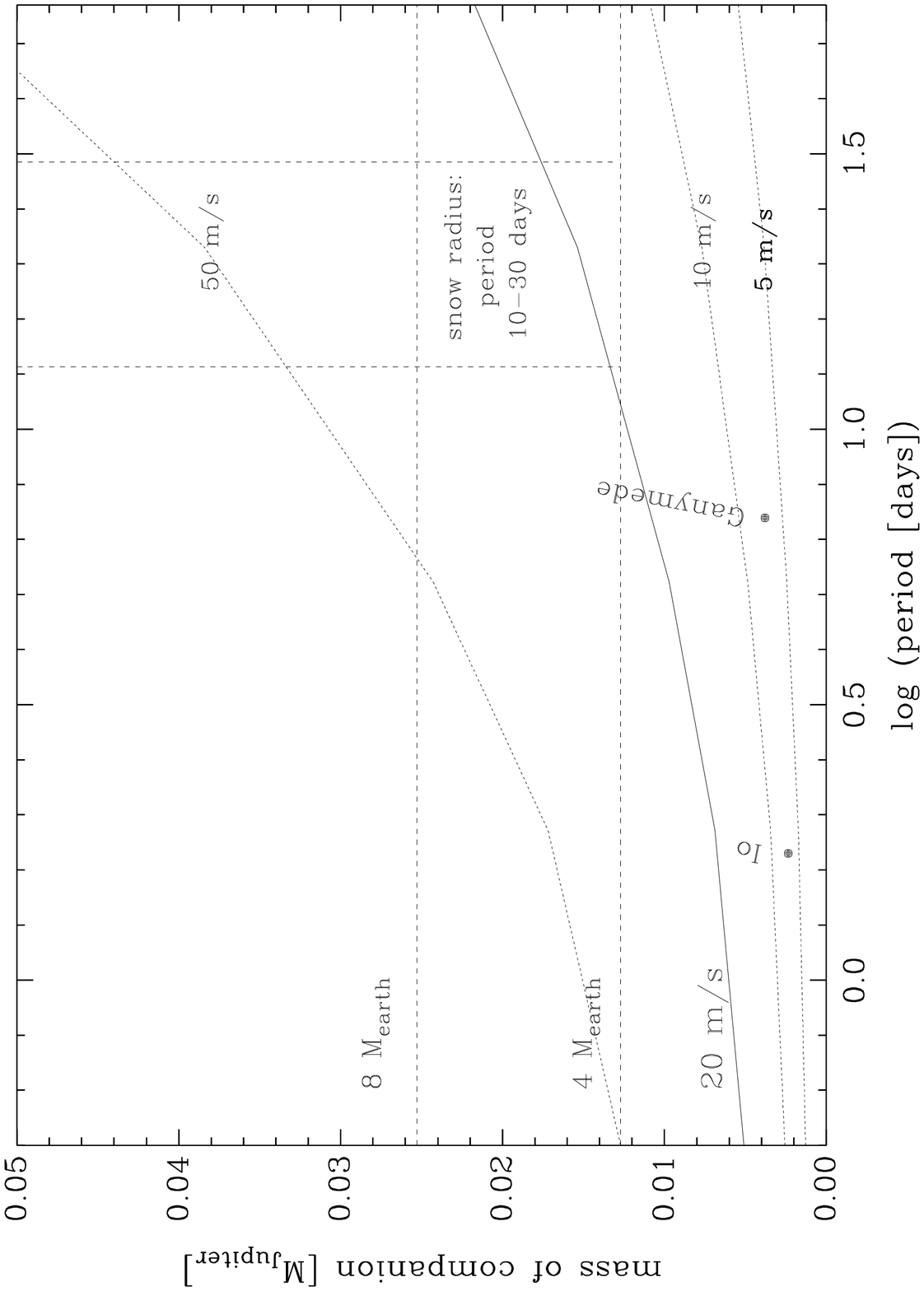}
\caption{The figure shows the semi-amplitude of the RV-variations caused
planets orbiting BDs. A BD of 40 $M_{Jupiter}$ is assumed. Planets of
only a few earth-masses could be detected if an accuracy of 20
$m\,s^{-1}$ is achieved. An accuracy of 5 $ms^{-1}$ is required in order
to detect objects analogue to Io and Ganymede. In the case of the disk
of a BD the ice-line would correspond to an orbital period of only 10 to
30 days.}
\end{center}
\label{bd_companions}
\end{figure}

What would be expected, if we go to stars of lower mass?  According to
Laughlin, Bodenheimer, \& Adams (2004), stars of lower masses had also
disk of lower masses and thus should also have planets of lower masses.
This idea is in good agreement with the observations, as Butler et
al. (2005) estimate that the fraction of Jupiter-mass planets of M-stars
is at least a factor of 5 smaller than of FGK-stars. However, up to
now it is not yet clear whether this is solely a result of the lower
masses of the proto-planetary disks, or at least partly due to
evaporation due to the strong UV radiation of young M-stars.

What would we expect, if we go to even lower masses, planets of brown
dwarfs (BDs) and very low mass stars (VLMSs)? For BDs the evaporation
due to the strong UV radiation is certainly irrelevant. If planets can
only form by core accretion, one would expect to find only planets of
very low mass ($\sim$ 0.01 $M_{Jupiter}$ or a few $M_{earth}$ in this
case).  Likewise, one may argue that BDs resemble Jupiter and we might
expect to see only Io or Ganymede-type objects. Again this implies that
BDs should have planets of only a few $M_{earth}$ (Desidera 1999).
However, even such planets could be detected with an accuracy of
RV-measurements of 5 to 10 $ms^{-1}$ (Fig.\,1).  On the other hand, one
may argue that BDs and VLMSs should have massive planets, simply because
there are many BD-BD binaries and the mass-ratio between a BD and a
massive planet is only $\sim$ 1:10. Since for field objects there is no
break in the initial mass function at 13 $M_{Jupiter}$, such ``BD-planet
binaries'' could be possible. Of course, the best evidence for the
existence of such objects is 2MASSW\,J1207334-393254 (Chauvin et
al. 2005). If BDs have massive planets, we would thus expect that there
is no correlation with metalicity, as these planets would not have been
formed by core-accretion.

Up to now, only two programs to search for planets by means of
RV-measurements have been carried out.  Joergens (2005) monitored 7
young BDs in the Chameleon cluster, and found one companion
candidate. Additionally, she found that the RV-variations caused by
activity decreased with the mass of the object. Guenther \& Wuchterl
(2003) monitored the 26 VLMSs and BDs, and fond apart from three
binaries one object which showed significant RV-variations:
LP944-20. Unfortunately, it is not yet clear, whether these are caused
by surface features, or by an orbiting planet.

A search program for planets of VLMSs and BDs with NAHUAL will thus
show, whether BDs and VLMSs have planets or not, and if so whether these
planets formed by core-accretion or not.

\subsection{Calibrating evolutionary tracks and the atmospheres of BDs}

The lack of knowledge of the true masses of VLMSs and BDs at young age
is a severe problem for this field of research. A dedicated search
program for eclipsing BD-BD, or BD-planet binaries should solve the
problem.  Eclipsing systems could best be found with NAHUAL in a survey
in which many BDs are observed but each BD is observed only three times.

NAHUAL will also allow to study the atmospheres of low-mass objects in
detail, because it is possible to obtain spectra covering simultaneously
J, H, and K-band at a resolution of $\lambda/ \Delta \lambda=50,000$.
Such observation would allow to determine $T_{eff}$, log(g) and the
abundances, for example.

Young BDs show clear signs of accretion. The presently available optical
data seems to indicate that $\dot M \sim M^2_*$ (Mohanty, Jayawardhana,
\& Basri 2005). Because the disk mass $M_d$ $\sim\,M_*$, one would
deduce that VLMSs and BDs should take much longer than solar-like stars
to form.  Since the flux of the Bracket $\gamma$ line is well correlated
with the accretion rate, observations with NAHUAL will shed more light
on to this question.

\subsection{Planets of active stars}

One problem of the RV-technique is that not only orbiting planets but
also stellar spots, plage regions, changes of the granulation pattern,
and oscillations can also lead to RV-variations. However, the only
effect that causes RV-variations that does not depend on wavelength is
an orbiting object. Thus, be carrying out RV-measurements at optical and
IR wavelength, it is possible to distinguish between orbiting planets
and other effects.

In interesting question is, whether the RV-scatter caused by stellar
activity becomes larger, or smaller when going from optical to IR
wavelength. Paulson et al. (2002) used their precise RV-data of the
Hyades stars in order to investigate the cause of the RV-scatter. They
find that the scatter is mainly caused by spots.  Plage regions are less
important.  This result is confirmed by RV-monitoring of the very active
star EK Dra (K\"onig et al. 2005). The main effect is that the 90 to
95\% light deficit of a spot causes a hump in the profile of a
spectral\-line which moves across it with the rotation of the star. We
modelled this effect and find that it is reduced by a factor of 10 at IR
wavelength, because the difference in brightness between a spot and the
photosphere is smaller in the IR.

\section{Scientific requirements}

In order to achieve the highest possible accuracy for the
RV-measurements, the top requirements are:

\begin{itemize}
\item Large spectral coverage: This requirement calls for
      a cross-dispersed Echelle spectrograph and a 2048x2048 
      HAWAII-2 PACE HgCdTe detector. NAHUAL will cover the
      whole region from 0.9 to 2.4 $\mu m$, with only some
      small gaps in the K-band. 
\item High signal-to-noise ratio: This requirement implies
      a high throughput and a big telescope, the GTC.
      Additionally, the instrument will be cooled to 70 K.
\item A resolution that is high enough to resolve the spectral lines:
      Given the $v\,sin\,i$-values of VLMSs and BDs, this implies
      a resolution of $\lambda/ \Delta \lambda \geq 50,000$.
\item An absorption cell for the wavelength-self reference.
\item AO-system to allow for a narrow slit, and
      for stabilising the star on the slit.
\item Stable environment: The instrument will be place at the
      Nasmyth platform, evacuated, and temperature stabilised.
      Kjelsen et al. (2005) has demonstrated that an amazing
      accuracy of 0.44 $m\,s^{-1}$ can be achieved with
      UVES, which is also placed at the Nasmyth platform,
      and also uses an absorption cell as a wavelength
      self-reference.
\end{itemize}

\begin{figure}[ht]
\vspace*{16.0cm}
\includegraphics{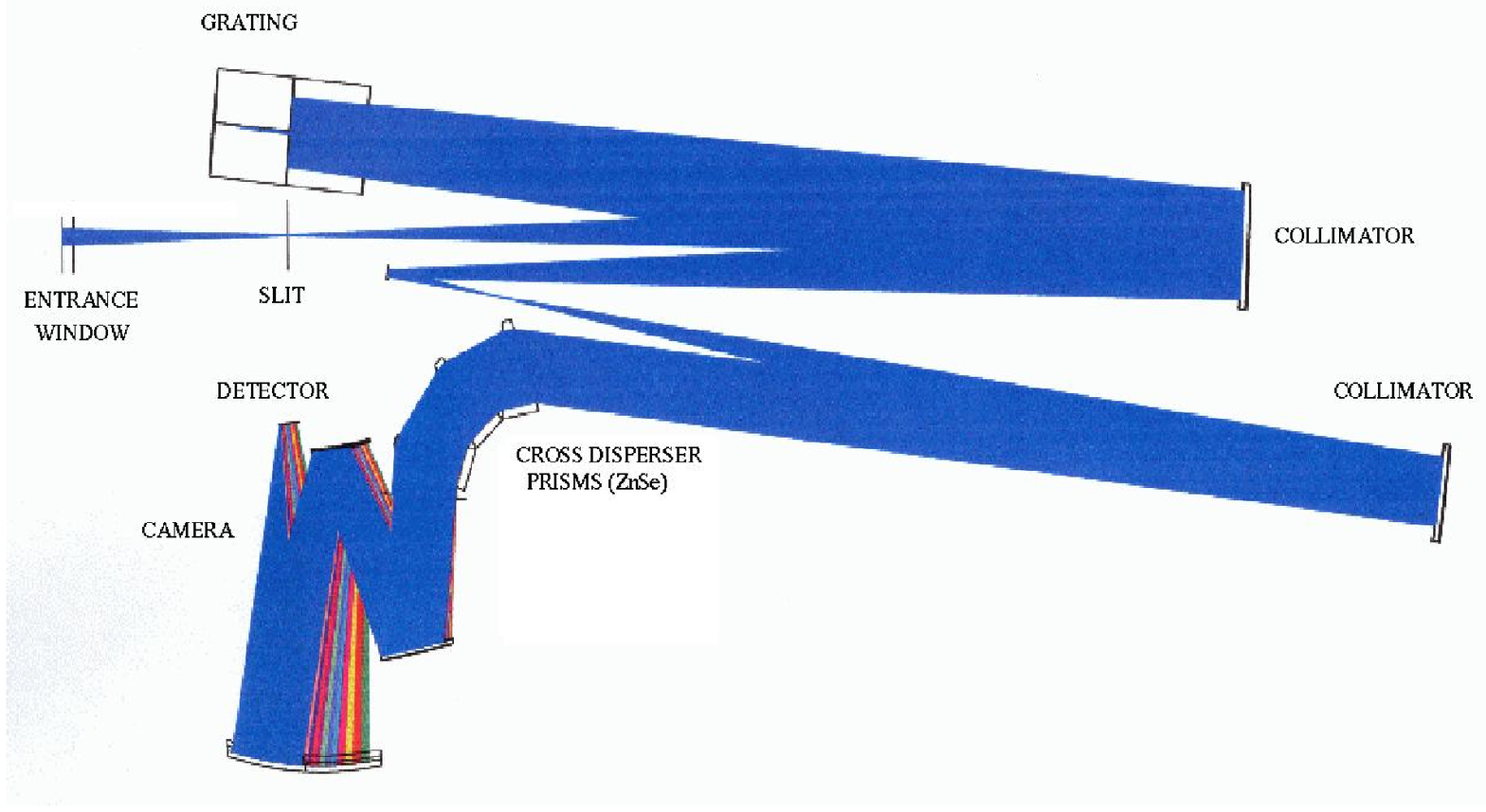}
\caption{The figure shows the conceptual optical design of NAHUAL} 
\label{design}
\end{figure}


\section{Conceptual design of NAHUAL}

The resolution of the spectrograph per arcsec is given by $ R\varphi =
2(d/D) tan \alpha_B$ where R is the resolution, $\varphi$ the slit width
on the sky, $d$ the diameter at the collimated beam, $D$ the telescope
diameter and $\alpha_B$ is the blaze angle.  The first thing to discuss
is whether a grating with $tan \alpha_B=4$, or 2 should be used. While a
$tan \alpha_B=4$-grating gives a higher resolution, it requires a
spectrograph camera of very short focal length with a large field of
view.  Unfortunately, we studied several possibilities but all of them
suffer from vignetting at the edge of the field. This essentially is
because a three mirror system has to be used. We thus have to choose a
$tan \alpha_B=2$-grating.  We also studied instrument concepts with
collimated beam diameters of 200 and 100 mm but finally decided for the
smaller beam diameter in order to keep the instrument manageable
(Fig.\,2). Forseen is a gold coated grating of 23.2 $gr\,mm^{-1}$ with
${\alpha_B}=63^o$.

These parameters already fix the slit-width to 0.175 arcsec for a
resolution of R=50000. NAHUAL will thus be work with an AO-system. A
natural seeing mode with an image-slicer is also being studied. By
placing a flat mirror in front of the Echelle grating NAHUAL will also
allow to take spectra with R $\sim$ 300, using only the three ZnSe
prisms of the cross-disperser.  Because the AO-system will not change
the f-ratio of the GTC, the focal length of the two collimators are
fixed to 1700 mm. Given the pixel size of the 2048x2048 HAWAII-2 PACE
detector, we designed a three-mirror camera of 420 mm focal length,
which has no vignetting even at the edge of the 40x40 mm field of
view. It is interesting to note that for designing the instrument we
used the GIANO (http://www.bo.astro.it/giano/) concept as a starting
point but then finally came up with a design that is closer to HARPS. In
the conceptual design NAHUAL is 2.5 m long and 1.2 m wide.

\section{The absorption cell laboratory}

In order to achieve a high accuracy of the RV-measurements an absorption
cell has to be placed in front of the spectrograph. Such a cell imposes
a large number of dark lines on to the spectrum of the object which are
recored simultaneously with the observations (self-reference
spectrograph). While an $I_2$-cell is commonly used in the optical
regime, the best choice of gases for an infrared cell still has to be
found. As part of the NAHUAL study, we have started a laboratory
experiment to try out various gas mixtures. The gases are mixed in a
controlled vacuum chamber and the spectra are measured with a
spectrophotometer which is available in the IAC optical laboratory.
Promising is a mixture of $N_2O$ (30\%), $H_2C_2$ (27\%), He (25\%), and
$CH_4$ (18\%).

\section{Schedule and Conclusions}

We have outlined the potential of NAHUAL for surveys of planets of VLMSs
and BDs. We would like to give one example for the power of NAHUAL at
the end. With UVES and the VLT it is possible to achieve an RV-accuracy
of 500 $ms^{-1}$ with an exposure time of 20 minutes for the BD LP944-20. With
NAHUAL it will be possible to achieve 5 $ms^{-1}$ with an exposure time of
only 5 minutes. Thus, we hope that NAHUAL will open up an entirely new
window for planet research. The plan is to finish the design study of
NAHUAL in 2006, and by 2010 to have the first light.

\acknowledgments{We would like to thank the whole NAHUAL team,
especially those that participated in the first and second NAHUAL meeting
in the paradores nationales in La Gomera (2004), as well as Segovia
(2005).  Funding for this work has been provided by the Spanish
Ministerio de Educaci\'on y Ciencia through Acci\'on Complementaria
AYA2004-22113-E.}

\end{document}